\documentclass[11pt]{emulateapj}
\usepackage{graphicx}
\usepackage{natbib}

\newcommand{\jj}[2]{\mbox{$J = #1\rightarrow#2$}}
\newcommand{\msunpc}{\mbox{M$_\odot$ pc$^{-2}$}}
\newcommand{\kms}{\mbox{km s$^{-1}$}}
\newcommand\cmv{\mbox{cm$^{-3}$}}

\begin{document}

\title{Evidence for Decay of Turbulence by MHD Shocks in Molecular Clouds via CO Emission}
\author{Rebecca L. Larson \altaffilmark{1}, Neal J. Evans II \altaffilmark{1}, Joel D. Green \altaffilmark{1,2}, \& Yao-Lun Yang \altaffilmark{1}} 
\altaffiltext{1}{University of Texas at Austin, Department of Astronomy, Austin, TX} 
\altaffiltext{2}{Space Telescope Science Institute, Baltimore, MD}

\begin{abstract}

We utilize observations of sub-millimeter rotational transitions of CO from a {\it Herschel} Cycle 2 open time program (``COPS'', PI: J. Green) to identify previously predicted turbulent dissipation by magnetohydrodynamic (MHD) shocks in molecular clouds. We find evidence of the shocks expected for dissipation of MHD turbulence in material not associated with any protostar. Two models fit about equally well: model 1 has a density of  10$^{3}$ cm$^{-3}$, a shock velocity of 3 \kms, and a magnetic field strength of 4 ${\mu}$G; model 2 has a density of 10$^{3.5}$ cm$^{-3}$, a shock velocity of $2$ km s$^{-1}$, and a magnetic field strength of 8 $\mu$G. Timescales for decay of turbulence in this region are comparable to crossing times. Transitions of CO up to $J$ of 8, observed  close to active sites of star formation, but not within outflows, can trace turbulent dissipation of shocks stirred by formation processes. Although the transitions are difficult to detect at individual positions, our {\it Herschel}-SPIRE survey of protostars provides a grid of spatially-distributed spectra within molecular clouds.  We averaged all spatial positions away from known outflows near seven protostars. We find significant agreement with predictions of models of turbulent dissipation in slightly denser (10$^{3.5}$ cm$^{-3}$) material with a stronger magnetic field (24 $\mu$G) than in the general molecular cloud. 

\end{abstract}

\maketitle

\section{Introduction\label{sec:Motivation}}

The dust in molecular clouds shields the interior from ultraviolet radiation, lowering the kinetic temperature to about 10 K. For this temperature, the isothermal sound speed is only about 0.2 km s$^{-1}$. In contrast, the observed linewidths of molecular transitions indicate much larger velocity dispersions. When these facts first became evident, \citet{1974ApJ...189..441G} suggested that the large linewidths were caused by overall collapse of the molecular cloud, rejecting the alternative of turbulent motions because such motions should decay on a timescale similar to a crossing time. \citet{1974ARA&A..12..279Z} noted that allowing all molecular clouds to collapse at free-fall would produce a star formation rate far higher than that of the Milky Way. \citet{1974ApJ...192L.149Z} suggested that the linewidths were indeed caused by turbulence, and that magnetic fields could slow the dissipation if the motions were sub-Alfv\'{e}nic. The idea that the linewidths are indicative of turbulent motions is widespread \citep{1999ApJ...522L.141M, 1987ApJ...319..730S, 1981MNRAS.194..809L}, but simulations \citep{1998ApJ...508L..99S, 2000A&A...356..287S} indicate that the turbulence should decay via shocks. Even when magnetic fields are included, simulations indicate that turbulence decays, but through magnetohydrodynamic (MHD) or C-shocks \citep{1971MNRAS.153..145M, 1980ApJ...241.1021D, 1993ARA&A..31..373D}, or through coherent, small-scale vortices \citep{2014A&A...570A..27G}. Strong evidence for decay has been found in diffuse and translucent clouds using chemical indicators \citep{2014A&A...570A..27G} and H$_2$ emission lines \citep{2005A&A...433..997F}. The latter method has also indicated evidence for decay on the outskirts of the Taurus molecular cloud \citep{2010ApJ...715.1370G}.
The idea that turbulence decays in denser molecular clouds is also widely accepted, but direct evidence for the shocks in opaque molecular clouds has proven elusive. The discovery of such evidence would resolve a long-standing question.

The sources of this turbulence are still somewhat unclear. One source for renewing the turbulence is outflows from forming stars. The outflows are hypersonic, but they may couple energy to the surrounding gas via magnetic disturbances. Thus, we may find evidence for enhanced dissipation in the surroundings of outflows, but we must avoid the outflows to avoid confusion. 

Much of the energy deposited in the gas by shocks is radiated away, and a likely route is through CO and H$_2$ rotational transitions, as well as through enhanced emission by atoms and ions, such as OI and CII. Attempts to find signatures of turbulent dissipation in lines of H$_2$, CII, and OI are frustrated by the fact that these lines could also be excited by low intensity photodissociation regions (PDRs) which cover the surface of all clouds \citep{2002ApJ...568..242L}. A renewed opportunity to detect the signature of the shocks has arisen due to the combination of new models, focusing on the mid-$J$ CO lines and the ability of the {\it Herschel} instruments (both HIFI and SPIRE) to detect these lines toward regions that are not directly affected by outflows.

Models from \citet{2012ApJ...748...25P} indicate that turbulent shocks enhance mid-$J$ CO emission over that from quiescent PDR models \citep{1999ApJ...527..795K}. Observational evidence for turbulent dissipation has been found in the Perseus B1-East 5 region by \citet{2014MNRAS.445.1508P}. They chose this region of the Perseus molecular cloud as one with no significant heating sources other than shocks, cosmic rays, and interstellar radiation fields (ISRF); there are no signs of embedded young stellar objects. Measurements of the CO \jj{5}{4} and \jj{6}{5} transitions show values enhanced over PDR models by \citet{1999ApJ...527..795K}. \citet{2014MNRAS.445.1508P} concluded that the filling factor of shocks toward this position was low ($0.15$\%), and that the timescale for turbulent dissipation was about a factor of 3 shorter than the flow crossing time \citep{2015MNRAS.447.3095P}. This is an important discovery, but it applies only to a single region and required a combination of data from different instruments. A confirmation with other techniques in different regions would add confidence that turbulent dissipation has finally been detected. Similar results have recently been obtained for apparently starless clumps within IRDCs by \citet{2015arXiv150300719P}.

In this work we present a sample of SPIRE spectroscopy of fields in molecular clouds including one near no protostar and seven near protostars.  Using portions of these fields unaffected by outflows or nearby protostellar radiation fields, we examine them for evidence of turbulent decay, as indicated by increased luminosity in mid-J CO lines.

\subsection{Object Selection\label{Objects}}

The objects that were analyzed were taken from the COPS-SPIRE sample (Green et al. in prep.) of low luminosity embedded protostars. These observations were not optimized for this project as they tend to have powerful outflows which produce emission in very high-$J$ lines. To avoid outflows, we selected objects with weaker outflows and with outflows well confined to certain regions, focusing on the regions away from the outflows. In addition, we obtained one region within a filament of the Taurus cloud that did not include any embedded source, thus providing a sample of the molecular cloud {\it without} any star formation. We refer to this observation as the isolated molecular cloud. 

Table \ref{sourcelist} describes the seven regions with embedded sources that were selected for this study. L1455-IRS3, IRAS03301-3111, and TMR1 were chosen because they can be easily spatially resolved in the SPIRE field. TMC1 and TMC1A have compact outflows that allow for analysis of the surrounding molecular cloud within the field of view. L1551-IRS5 has a large, but well collimated, bipolar outflow in the North-East/South-West direction. L1157 has distinct collimated bipolar outflows in the North-South direction. \\

\section{Observations and Reduction\label{sec:Procedure}}

Data sets were collected by the {\it Herschel Space Observatory}, a 3.5-m space telescope \citep{2010A&A...518L...1P}, using the SPIRE (Spectral and Photometric Imaging Receiver) Fourier Transform Spectrometer \citep{2010A&A...518L...3G}. The SPIRE short-wavelength (SSW) spectrum covers the 194-313 $\mu$m range over an array of 37 pixels (two of which were not functioning) that project beams at 33$\arcsec$ separation. The SPIRE long-wavelength (SLW) spectrum covers the 303-671 $\mu$m range over an array of 19 pixels, and these beams are separated by 51$\arcsec$. For this work, we restrict the analysis to CO lines within only the SLW data sets, which cover CO transitions up to \jj{9}{8}. We also examined the SSW pixels for higher-$J$ lines, but most of our sources did not show evidence for CO in the SSW range, and the higher-$J$ CO lines we could identify could not be resolved from the outflows of the forming stars.\\

\begin{deluxetable}{l c c c c }
\tabletypesize{\scriptsize}
\tablecaption{Source List \label{sourcelist}}
\tablewidth{0pt}
\tablehead{\colhead{Source} & \colhead{Cloud} & \colhead{Dist} & \colhead{RA (J2000)} & \colhead{Dec}}
\startdata
L1455-IRS3 & Per & 250 & 03h28m00.4s & +30d08m01.3s \\
IRAS 03301+3111 &  Per & 250 & 03h33m12.8s & +31d21m24.2s \\
L1551-IRS5 & Tau & 140 & 04h31m34.1s & +18d08m04.9s \\
TMR 1 & Tau & 140 & 04h39m13.9s & +25d53m20.6s \\ 
TMC 1A & Tau & 140 & 04h39m35.0s & +25d41m45.5s \\
TMC 1 & Tau & 140 & 04h41m12.7s & +25d46m35.9s \\
L1157 &  Core & 325 & 20h39m06.3s & +68d02m16.0s \\
\hline 
Isolated MC & Tau & 140 & 04h39m53.0s & +25d45m00.0s
\enddata
\tablecomments{ ~List of protostellar sources and the isolated molecular cloud position used in this sample, sorted by RA. Distances are measured in parsecs.}
\end{deluxetable}

\begin{figure}[ht]
\center
\includegraphics[width=0.5\textwidth]{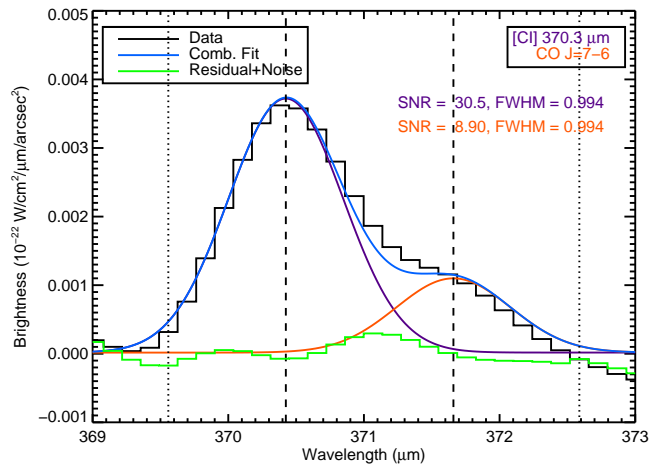}
\caption{Example of line fit for the blended CO \jj{7}{6} and [C I] $^{3}P_2 \rightarrow~ ^{3}P_1$ lines from the combined average of the non-outflow pixels from Figure \ref{combnooutflow}.  
}
\label{co76nooutflow}
\end{figure}

\indent	We separate each data set by pixel, including the object itself and the area around it. The continuum in each pixel was first calculated using an automated routine written in Interactive Data Language (IDL) and was then subtracted from the original spectrum to create a flat (continuum-subtracted) spectrum. The flat spectra from each pixel in the combination were averaged together to improve the signal-to-noise ratio. The CO lines in each average were fitted by a Gaussian function which yields the intensity integrated over the line. This intensity was converted to the units used in the models (erg s$^{-1}$ cm$^{-2}$ arcsec$^{-2}$). Upper limits for the non-detections were calculated by taking the noise level and multiplying by 3 to get the 3${\sigma}$ value. The CO \jj{7}{6} line flux estimate is particularly uncertain as it is blended with the [C I] line which is very prominent in the isolated molecular cloud field.
Figure \ref{co76nooutflow} shows an example of the blended CO \jj{7}{6} and [C I] $^{3}P_2 \rightarrow~ ^{3}P_1$ lines and an overlay of the double-Gaussian fit that was applied to distinguish the fluxes from each line individually.  \\

\subsection{Pixel Combinations\label{sec:Pixels}}

For the isolated molecular cloud, we could average the spectra over all the pixels. For the embedded sources, we used the following combinations of pixels:
    
\begin{itemize}

\item Center - The central pixel (labeled SLWC3), containing the protostar. The SPIRE beam size is large enough that the continuum source is point-like and restricted to the center pixel.

\item Outflows - These pixels contain outflows driven by the protostar. For each observation, we compared the SPIRE pixels to ground-based CO \jj{2}{1} and \jj{1}{0} contour maps from Kang et al. (submitted).  An example of this overlay and comparison can be found in Figure \ref{L1157Contour}.

\item Non-outflows -  This category includes all pixels not ``contaminated'' by the protostar or outflow emission. These pixels should be the best representation of the gas around a forming protostar, but not directly affected by the outflow.

\begin{figure}[ht]      
\center
\includegraphics[width=0.45\textwidth]{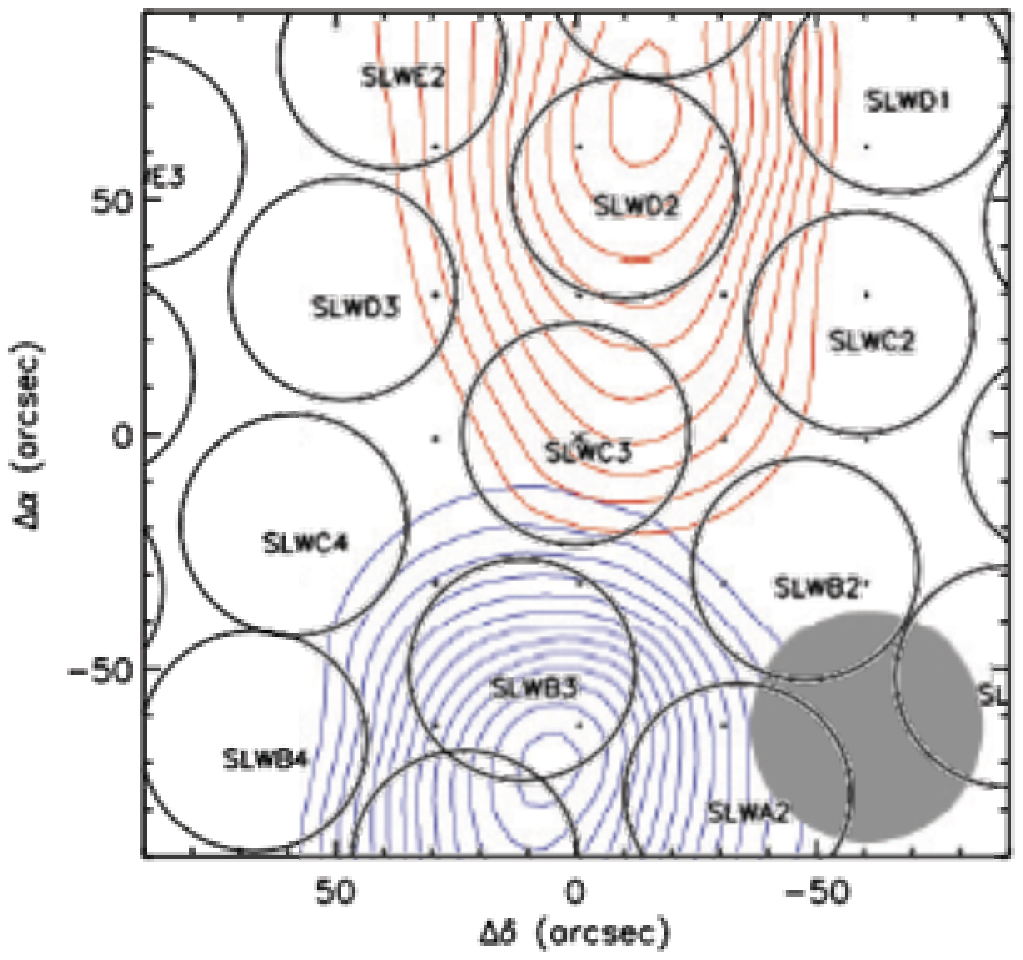}
\includegraphics[width=0.45\textwidth]{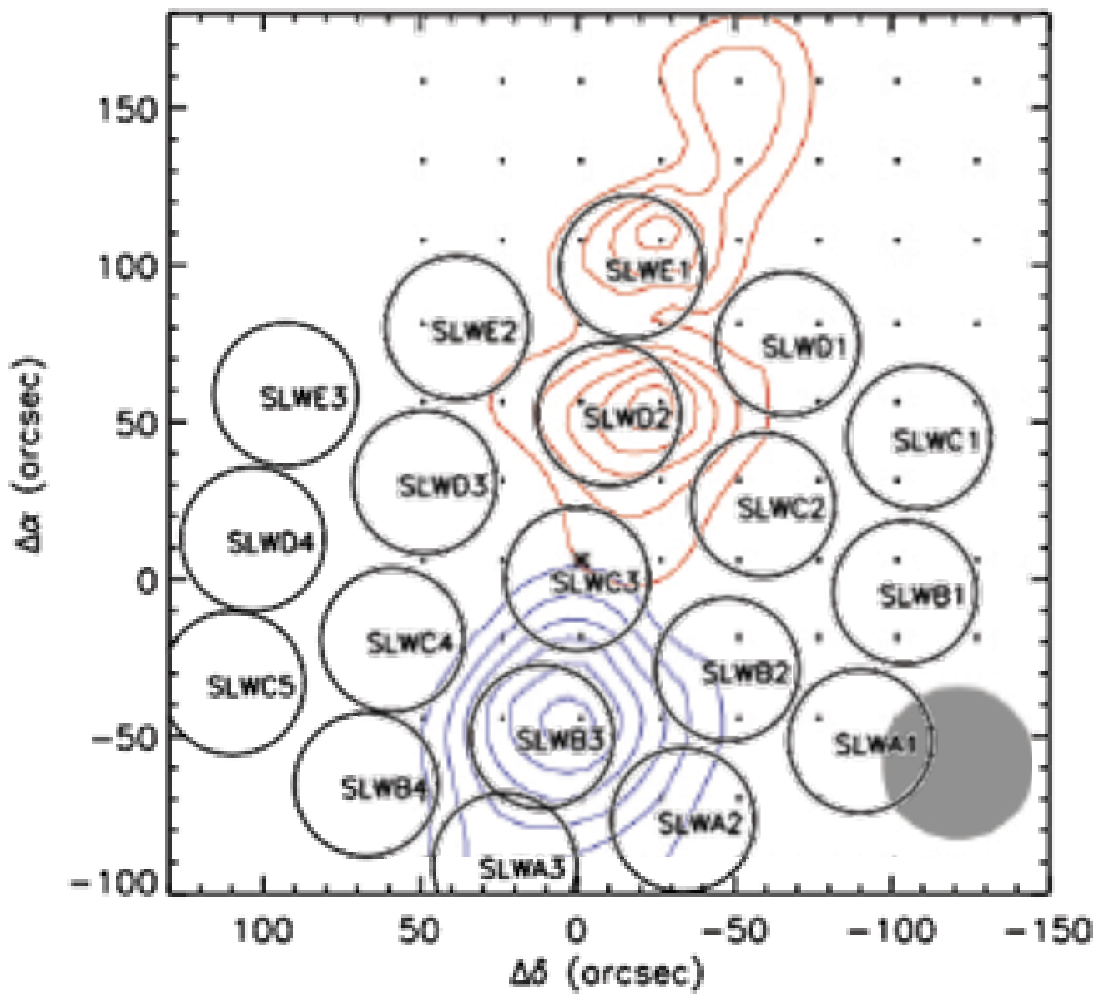}
\caption{Example of outflow map from object L1157. {\bf Top:} An overlay of the pixels of the SLW array from the {\it Herschel} telescope over the $^{12}$CO \jj{1}{0} contour map of the outflows from the object made by Kang et al. (submitted). The velocity ranges used for the blue and red components are $-3.8$ km s$^{-1}$ to 4.0 km s$^{-1}$ and 9.3 km s$^{-1}$ to 10.6 km s$^{-1}$, respectively. {\bf Bottom:} The SLW pixels overlaid on the $^{12}$CO \jj{2}{1} contour map. The velocity ranges used for the blue and red components are $-0.6$ km s$^{-1}$ to 4.3 km s$^{-1}$ and 8.6 km s$^{-1}$ to 9.4 km s$^{-1}$, respectively.}
\label{L1157Contour}
\end{figure}

\end{itemize}

\section{Analysis\label{sec:Analysis}}

We compare our data to three shock models supplied by A. Pon (pers. comm.). \citet{2012ApJ...748...25P} modeled integrated intensities of various CO rotational transitions in units of erg s$^{-1}$ cm$^{-2}$ arcsec$^{-2}$, considering three values for density:  $10^{2.5}$ cm$^{-3}$, $10^{3.0}$ cm$^{-3}$, and $10^{3.5}$ cm$^{-3}$. These values are based on the molecular cloud chemical composition originally used in \citet{1996ApJ...456..611K, 1996ApJ...456..250K}. They considered shock velocities of 2 and 3 km s$^{-1}$ for a cloud temperature of 10 K, corresponding to Mach 12 and Mach 17 respectively \citep{2012ApJ...748...25P}. These speeds are consistent with turbulence in molecular clouds but are significantly lower than the speeds of protostellar outflows. Magnetic field strength (shown here in ${\mu}$G) in a molecular cloud is defined by $$ B = bn_{H}^{k}$$ where $n_H$ is the number density of hydrogen nuclei and $b$ and $k$ are fitting parameters. The value of $k=0.5$ corresponds to a constant ratio of magnetic energy density to thermal energy density \citep{2007ARA&A..45..565M} and is the value assumed in these models. The strength of the magnetic field parallel to the shock front ranges from 3 ${\mu}$G to 24 ${\mu}$G, characterized by $b=0.1$ and $b=0.3$. The shock thickness is smaller and the energy released in line radiation is larger for a weaker field. Therefore, these values were chosen as they are slightly lower than, but still consistent with, those determined by \citet{2010ApJ...725..466C}. 

We also compare our data to CO predictions from  \citet{2012ApJ...748...25P}, which are based on the PDR models of \citet{1999ApJ...527..795K}. These PDR models have interstellar radiation field (ISRF) values specified in units of Habings ($1.6 \times 10^{-3}$ erg cm$^{-2}$ s$^{-1}$). 

These models also assume line widths of 1.5 km s$^{-1}$, which are on the scale of the shock model velocities, and a total $A_{\rm V}$ of 10. They do not take into account CO freeze-out and also assume plane parallel geometry. The densities used for the PDR models correspond to the densities of the shock models and all assume a G$_{0}$ value of 0.31, which is reasonably consistent with actual values for Taurus. \citet{2009ApJ...701.1450F} found a value of 0.2 Habings.

\begin{figure}[h!]  
\center
\includegraphics[width=0.45\textwidth]{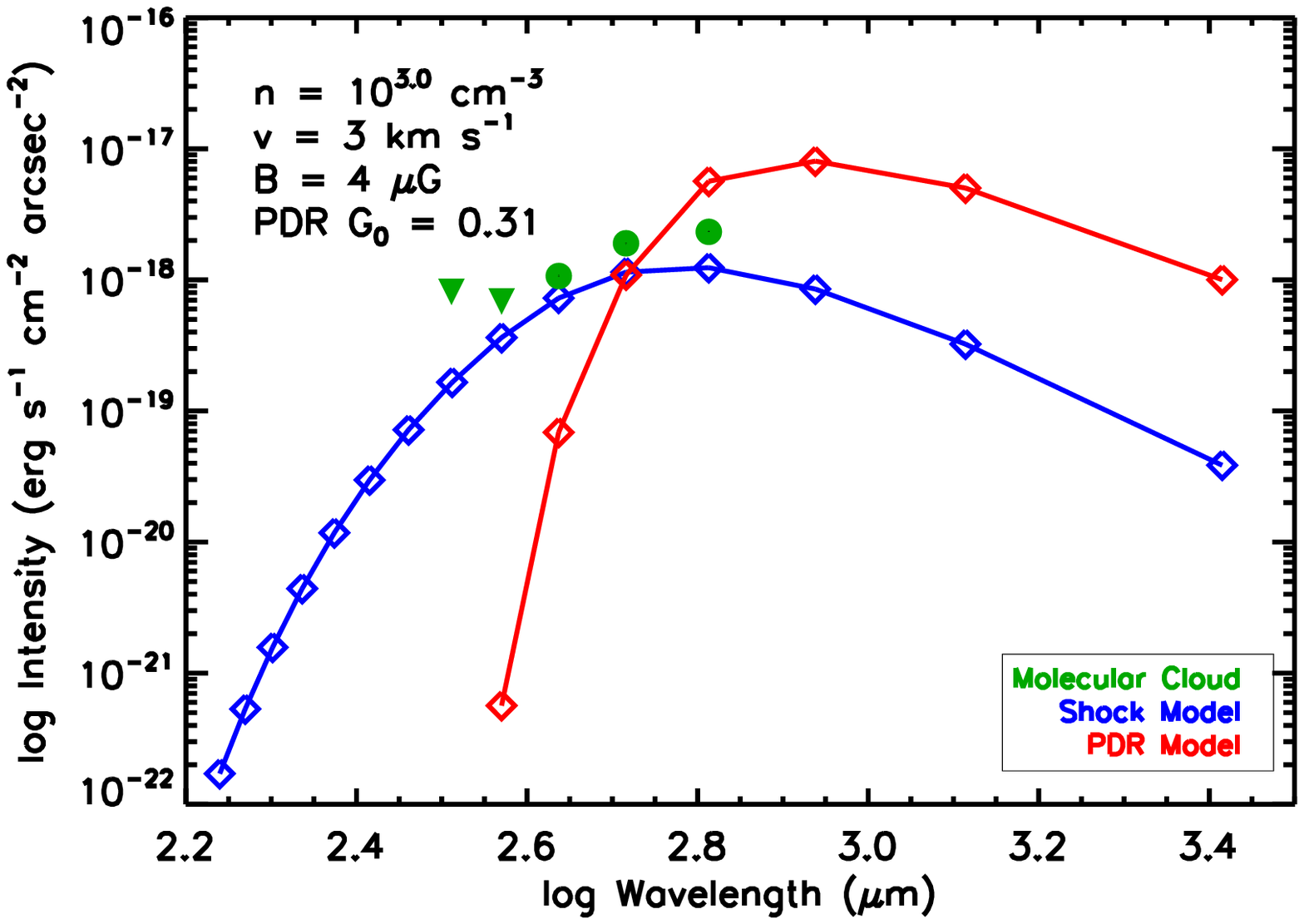}
\includegraphics[width=0.45\textwidth]{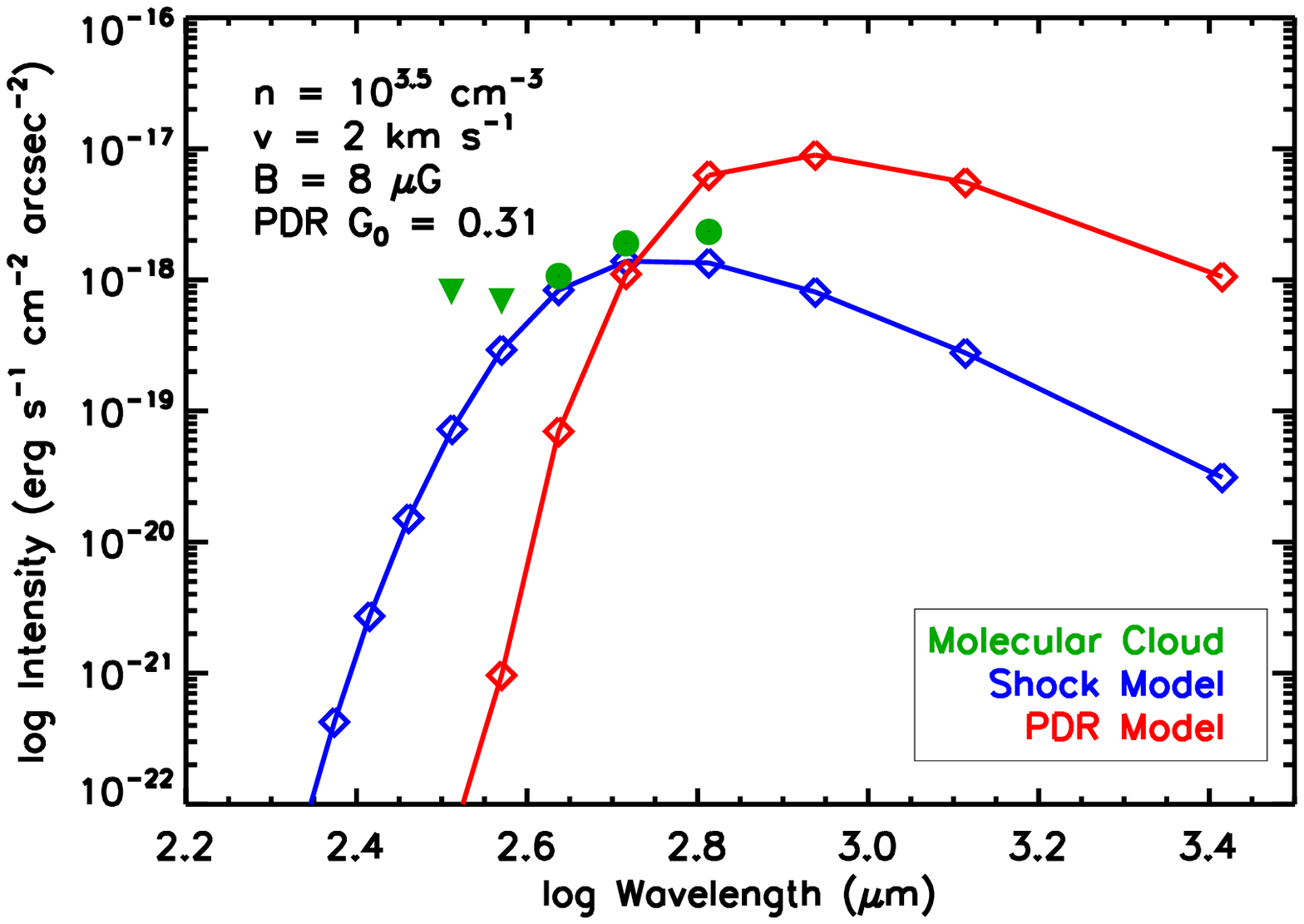}
\caption{Isolated Molecular Cloud: Observed CO emission lines from CO \jj{8}{7} (left) to CO \jj{4}{3} (right) from all pixels (green) plotted over two of the models they best fit. Triangles are estimated 3${\sigma}$ upper limits for non-detections. PDR models use a G$_{\rm 0}$ value of 0.31. Error bars are present but might be too small to distinguish. {\bf Top:} Intermediate density ($n = 10^{3.0}$ cm$^{-3}$), highest velocity ($v = 3$ km s$^{-1}$) and lower magnetic field strength (B = 4 ${\mu}$G). {\bf Bottom:} Highest density ($n = 10^{3.5}$ cm$^{-3}$), lower velocity ($v = 2$ km s$^{-1}$) and lower magnetic field 
strength (B = 8 ${\mu}$G). }
\label{L1527n30v3b1}
\end{figure}

\begin{figure}[h!]  
\center
\includegraphics[width=0.45\textwidth]{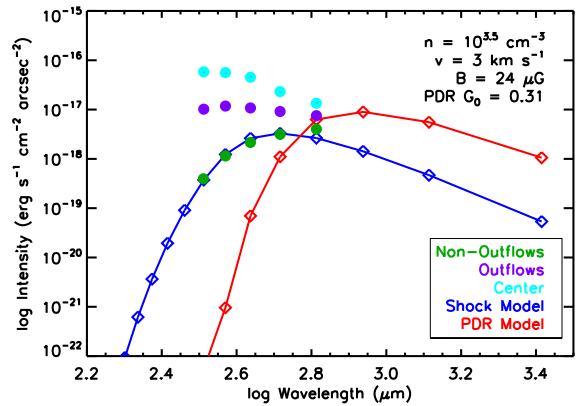}
\caption{Grand average of all seven objects: Observed CO emission lines from CO \jj{8}{7} (left) to CO \jj{4}{3} (right) from the center (light blue), outflow (purple), and non-outflow (green) pixels plotted over the model they best fit to: highest density ($n = 10^{3.5}$ cm$^{-3}$), highest velocity ($v = 3$ km s$^{-1}$), and higher magnetic field strength (B = 24 ${\mu}$G). PDR model uses a $G_{\rm 0}$ value of 0.31. Error bars are present but might be too small to distinguish.}
\label{Combinedn35v3b3}
\end{figure}

We present these models in Figure \ref{L1527n30v3b1} (isolated molecular cloud) and Figure \ref{Combinedn35v3b3} (grand average). The models show lines ranging from \jj10\ on the right to higher $J$ levels toward the left.  The model parameters are indicated in each figure panel. The plotted line fluxes are those taken from the averages obtained in the Center, Outflows, and Non-outflows composites: green (non-outflow pixels), purple (outflow pixels), and light blue (central protostar). For the isolated molecular cloud, all pixels are considered ``non-outflow'' (green).  Upper limits to fluxes are indicated by an upside-down triangle of the appropriate color.  The red points/curve show the intensities predicted for PDR models, while the dark blue points/curve are for the MHD shock model.  These are clearly distinguished in the models in that the shock spectra dominate over the contribution from PDRs for transitions from \jj54\ and higher \citep{2012ApJ...748...25P}.

\subsection{Isolated Molecular Cloud\label{L1527}}

We begin with an isolated portion of the Taurus molecular cloud with no embedded protostar. It is centered on a filament, far from any apparent source.  The emission is weak (we do not detect transitions above \jj{6}{5}) but the higher-$J$ lines are clearly stronger relative to the lower-$J$ lines than expected for a PDR model. Their strengths are reasonably consistent with both the $10^{3.0}$ cm$^{-3}$ density, highest velocity ($v = 3$ km s$^{-1}$), and lower magnetic field strength (B = 4 ${\mu}$G) model and the highest density ($n = 10^{3.5}$ cm$^{-3}$), lower velocity ($v = 2$ km s$^{-1}$) and lower magnetic field strength (B = 8 ${\mu}$G) model of \citet{2012ApJ...748...25P}, as shown in Figure \ref{L1527n30v3b1}. Improved sensitivity to detect higher-$J$ lines is required to further constrain the higher-$J$ lines and constrain the choice of model. The fact that even the low $G_0$ PDR models overestimate the observations of the \jj43\ line indicates that the PDR contribution to the higher $J$ lines would be even less than shown in Figure 3. Even so, the observations exceed by an order of magnitude the 
PDR predictions for \jj65. \citet{2014MNRAS.445.1508P} considered
other PDR models from different groups and found that they did not substantially
change their conclusions that shocks were required to match the observations of the higher $J$ lines. However, PDR models that produce substantially higher
excitation of the higher $J$ CO lines could conceivably reproduce the observations.

\subsubsection{Dissipation Timescale}

The fact that the observations match the model for $J$ values above \jj{5}{4} allows us to estimate the dissipation timescale for the turbulence at the position of the isolated molecular cloud. We treat the column through the cloud defined by a single pixel of the SPIRE array as typical of any region in the cloud which does not contain a protostar. The turbulent energy in this column is given by
\begin{equation}
E_{\rm t} = 0.5 \Sigma_{\rm m} \sigma^2 \Omega_{\rm b} d^2
\end{equation}
where $\Sigma_{\rm m}$ is the mass surface density, $\sigma$ is the 3-D rms turbulent velocity dispersion, $\Omega_{\rm b}$ is the beam solid angle in sr, and $d$ is the distance to the cloud. The rate of energy dissipation through shocks is
\begin{equation}
L_{\rm sh} = 4 \pi d^2 F_{\rm CO} f_{\rm CO}^{-1}
\end{equation}
where $F_{\rm CO}$ is the total observed line flux in all the CO lines and $f_{\rm CO}$ is the fraction of the total line emission emitted in CO. The flux in CO is the sum of all the intensities in
the best fitting model times the solid angle, $\Omega_b$, ($F_{\rm CO} = I_{\rm CO} \Omega_{\rm b} $). The timescale becomes
\begin{equation}
t_{\rm d} = E_{\rm t}/L_{\rm sh} = \frac{f_{\rm CO} \Sigma_{\rm m} \sigma^2}{8\pi I_{\rm CO}} 
\end{equation}
where $I_{\rm CO}$ is the sum of all the intensities of CO lines from the best fitting shock model in units of erg s$^{-1}$ cm$^{-2}$ sr$^{-1}$. The shock velocity in the models of \citet{2012ApJ...748...25P} is expressed in terms of the one-dimensional velocity dispersion ($\sigma_{\rm 1D}$) as $v_{\rm sh} = 3.2 \sigma_{\rm 1D} = (3.2/\sqrt{3})\sigma$, so $\sigma = 0.54 v_{\rm sh}$ For the two best fitting models, $v_{\rm sh} = 2$ or 3 \kms, $n = 10^{3.5}$  or $10^{3.0}$ \cmv, and $I_{\rm CO} = 2.11$ or $2.16 \times 10^{-7}$ erg s$^{-1}$ cm$^{-2}$ sr$^{-1}$.

We determined $\Sigma_{\rm m}$ from maps of extinction based on stellar reddening from \citet{2011ApJ...737..103S} and far-infrared emission from \citet{1998ApJ...500..525S} which yielded estimates of $E(B-V)$ of $2.49\pm 0.02$ mag and $2.89\pm 0.02$ mag respectively\footnote{http://irsa.ipac.caltech.edu/applications/DUST/}, where the uncertainties clearly do not include systematics. We convert to $A_{\rm V}$ using $R_{\rm V} = 5.5$, found to best match general molecular cloud extinction laws \citep{2009ApJ...690..496C}. Then we convert to $\Sigma_{\rm m}$ using the extinction efficiency from \citet{2001ApJ...548..296W} \footnote{http://www.astro.princeton.edu/~draine/dust/dustmix.html} for $R_{\rm V} = 5.5$, Case A (newer models). The results are 150 \msunpc\ \citep{2011ApJ...737..103S} or 175 \msunpc\ \citep{1998ApJ...500..525S}. If $R_V = 5.5$, Case B (older model) grains are used, the values are 206 and 240 \msunpc. For simplicity, we assume $\Sigma_{\rm m} = 150$ \msunpc\ in what follows, providing a lower limit
to the dissipation timescale.

For the model with $v_{\rm sh} = 2$, $ t_{\rm d} = 1.6$ Myr;
in the model with $v_{\rm sh} = 3$, $ t_{\rm d} = 3.3$ Myr.
Following the procedure in
Pon et al. 2012
we use the linewidth-size relation from 
\citet{1987ApJ...319..730S},
$ r({\rm pc}) = (\sigma_{\rm 1D}/0.72 \kms)^2$ where $\sigma_{\rm 1D} = \sigma/\sqrt{3}$
to estimate a cloud radius of 0.77 pc for the $v_{\rm sh} = 2$ \kms\ model.
Still following 
Pon et al. 2012
we estimate a flow-crossing time from $t_{\rm cr} = 2R/\sigma_{\rm 1D}$ of 2.3 Myr for the $v_{\rm sh} = 2$ \kms\ model and 5.1 Myr for the $v_{\rm sh} = 3$ \kms\ model. The ratio of $t_{\rm d}$ to 
$t_{\rm cr}$ is thus 0.94 or 0.65 for the two models. Larger values of $\Sigma_{\rm m}$ would increase these ratios by a factor up to 1.6.
This analysis shows that the turbulence  dissipation timescale derived from observations and best-fit model is comparable to the flow-crossing timescale. 
\citet{2014MNRAS.445.1508P}
found a ratio of 1/3 (originally published as 3, but corrected in an Erratum)
\citep{2015MNRAS.447.3095P}
in the Perseus cloud.

\subsection{Embedded Sources\label{embedded}}

The fields around the embedded sources can provide a sample of denser cores, where turbulence may be enhanced. The challenge is to avoid mixing in positions toward outflows, which produce high velocity shocks and strong emission from high-$J$ levels of CO. An example of a field around an embedded protostar (L1157) is shown in Figure \ref{L1157Contour}. L1157 provides a particularly strong test case, as the outflows (derived from CO \jj{1}{0} and \jj{2}{1} maps by Kang et al., submitted) are spatially well-defined in the pixel array and separation of outflows and non-outflow pixels is easily performed. For this case the pixels that do not have any influence from the protostellar object, and which are included in the non-outflows combination, are A1, B1, C1, C5, D4, and E3 of the SLW pixel array. Note that in some cases where a CO \jj{1}{0} contour map was unavailable, the SPIRE field of view exceeded that of our comparison CO map; for these cases, there is some possible contamination by larger scale outflows. The line fluxes were then compared to the models produced by \citet{2012ApJ...748...25P} to determine if the CO lines do follow the decay of turbulence trend, and to which properties each is most closely correlated.

\subsection{Grand Average Spectra}

\begin{figure}[ht]
\center
\includegraphics[width=0.45\textwidth]{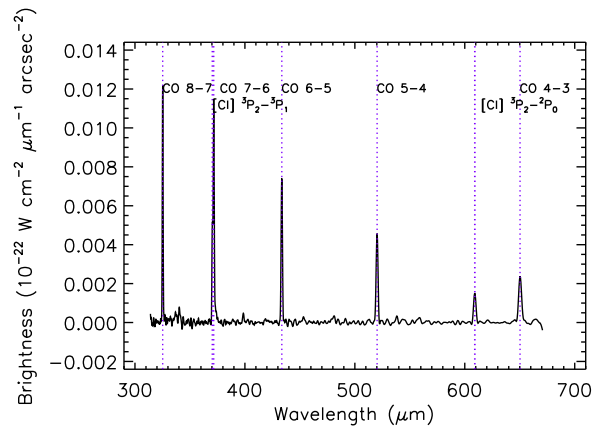}
\caption{The flat spectrum averaged over all pixels encompassing the protostellar outflows for all seven sources in this study.}
\label{comboutflow}
\end{figure}

\begin{figure}[ht]
\center
\includegraphics[width =0.45\textwidth]{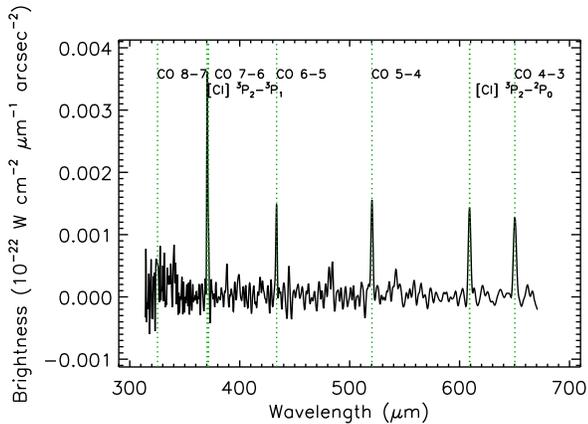}
\caption{The flat spectrum averaged over all pixels {\bf not} contaminated by protostellar outflows for all seven sources in this study.}
\label{combnooutflow}
\end{figure}

When we compared the observational data to the models, it was clear that none of the objects follow the PDR model. The data are a good match to the \citet{2012ApJ...748...25P} shock models, specifically to those of the highest (10$^{3.5}$ cm$^{-3}$) density, highest velocity ($v = 3$ km s$^{-1}$) and higher magnetic field strength (B = 24 ${\mu}$G). However, the only field that shows a clear difference between the outflow and the non-outflow pixels is around L1157. This could be because the object is oriented such that the pixels that contain outflows are clearly distinguishable from those that do not (see Figure \ref{L1157Contour}); we may still be suffering from confusion with outflows in other fields. There could be additional influence by the outflows in the pixels which have been chosen as non-outflow pixels and this may account for the close correlation of the fluxes in the outflow and non-outflow pixels around all but L1157. In order to improve the signal-to-noise ratio, we computed a grand average of all seven objects to search for general trends. Spectra for all center pixels, outflows, and non-outflows were (separately) averaged together and compared with the \citet{2012ApJ...748...25P} models (Figure \ref{Combinedn35v3b3}). The CO \jj{8}{7} through \jj{4}{3} line fluxes derived from pixels classified as outflow, non-outflow, and central protostar are indicated by purple, green, and light blue points respectively.  The grand average spectra for the outflow and non-outflow positions are shown in Figure \ref{comboutflow} and Figure \ref{combnooutflow} respectively. The grand average spectra show clear distinctions between outflow and non-outflow pixels once the
confusing line from [C I] at 370 \micron\ has been removed, as seen in Figure 1.

These grand average line intensities are fitted by the highest density, highest velocity and higher magnetic field model from \citet{2012ApJ...748...25P}. The center pixels, outflows, and non-outflows all follow a distinctly different trend and the grand average of the non-outflow pixels follows the shock models very closely. (See Figure \ref{Combinedn35v3b3}). While there is still a concern about contamination from the outflows themselves, the different patterns for the distribution of emission over rotational levels suggests that the non-outflow positions are tracing dissipation of low velocity shocks rather than the high-velocity shocks in the outflow and center positions. If so, the fact that a model with higher density, higher velocity, and higher magnetic field  compared to the isolated molecular cloud position fits these data suggests that the infalling envelopes can be stirred by infall or by propagation of the outflow energy via magnetic fields. However, because the grand average spectrum is a mixture of data from different clouds, calculation of a quantitative dissipation timescale would be inappropriate.

\section{Conclusions\label{sec:Conclusion}}

We observe evidence of turbulent shocks in a region of the Taurus cloud far from protostellar sources. This result supports the conclusions of \citet{2014MNRAS.445.1508P}, using a different technique in a different cloud. Together with similar conclusions in regions of more diffuse gas 
\citep{2014A&A...570A..27G} and \citep{2005A&A...433..997F} 
and the results of
\citet{2014MNRAS.445.1508P,2015MNRAS.447.3095P,2015arXiv150300719P} in gas
of similar density,
our data provide strong evidence for the dissipation of turbulence in molecular clouds. While substantial uncertainties remain in the timescale for dissipation,
the data support simulations in which MHD turbulence decays within a crossing 
time. Particularly for a cloud like Taurus, with relatively distributed and low
star formation activity, the question of what inputs balance the decay presents
an interesting challenge to theorists.

The measured intensities for the grand average spectra of embedded protostars show clear differences between the average of positions in the outflow and positions not in the outflow. These results suggest that we are not simply seeing contamination from the outflow shocks. Instead, we tentatively identify the CO spectra toward the non-outflow positions as a good match to models of turbulent decay in dense gas. Since these spectra are all in dense cores surrounding forming stars, the fact that the best fitting models have higher densities is not surprising.

\acknowledgements

This work was supported by NSF Grant AST-1109116. R.L. acknowledges support from the Karl G. Henize Endowed Scholarship, the John W. Cox Endowment for the Advanced Studies in Astronomy, and the College of Natural Sciences Summer Undergraduate Research Fellowship at the University of Texas at Austin.  R.L. would also like to thank Dustin Davis for technical support. The authors thank Andy Pon and Doug Johnstone for providing comparison models and helpful comments, and the anonymous referee for insightful comments. J.G. and Y.-L. Y. acknowledge support from NASA Herschel Science Center Cycle 2 grants.  This research has made use of NASA's Astrophysics Data System, and the SIMBAD database operated at CDS, Strasbourg, France. 

\bibliographystyle{apj}

\bibliography{references}

\end{document}